*Chapter V.3 :*

# *Coil creep and skew-quadrupole field components in the Tevatron*

*Gerald Annala, David Harding, Mike Syphers*

*Fermi National Accelerator Laboratory*

During the start-up of Run II of the Tevatron Collider program, several issues surfaced which were not present, or not seen as detrimental, during Run I. These included the repeated deterioration of the closed orbit requiring orbit smoothing every two weeks or so, the inability to correct the closed orbit to desired positions due to various correctors running at maximum limits, regions of systematically strong vertical dipole corrections, and the identification of very strong coupling between the two transverse degrees-of-freedom. It became apparent that many of the problems being experienced operationally were connected to a deterioration of the main dipole magnet alignment, and remedial actions were undertaken [1]. However, the alignment alone was not enough to explain the corrector strengths required to handle transverse coupling.

With one exception, strong coupling had generally not been an issue in the Tevatron during Run I. Based on experience with the Main Ring, the Tevatron was designed with a very strong skew quadrupole circuit to compensate any quadrupole alignment and skew quadrupole field errors that might present themselves. The circuit was composed of 48 correctors placed evenly throughout the arcs, 8 per sector, evenly placed in every other cell. Other smaller circuits were installed but not initially needed or commissioned. These smaller circuits were composed of individual skew quadrupole correctors on either side of the long straight sections. These circuits were tuned by first bringing the horizontal and vertical tunes near each other. The skew quadrupoles were then adjusted to minimize tune split, usually to less than 0.003. Initially, the main skew quad circuit (designated T:SQ) could accomplish this global decoupling with only 4% of its possible current, and the smaller circuits were not required at all. The start-up of Run Ib was complicated by what was later discovered to be a rolled triplet quadrupole magnet in one of the Interaction Regions [2]. This led to a reduction in luminosity of nearly 50%, as well as operational confusion until it was uncovered.

By the time Collider Run II began, the current needed on the main SQ circuit had increased to 60% of its maximum value. Some of the smaller circuits were also needed to fully decouple the tunes. With this history, several studies were performed early in Run II to search for strong local coupling sources like the triplet quadrupole, but without success. The strong corrector settings were indicative of a much larger problem than a single rolled magnet, and the locality of the error was hard to deduce from the setting of a global correction system. Several possible reasons for the increase in coupling were investigated.

## STRONG SYSTEMATIC STEERING CORRECTION

In late 2002 regions of the Tevatron were found to contain vertical steering magnets whose average strength was required to be non-zero in order to produce a smooth trajectory as seen on the Beam Position Monitors. Compared to the 0.7 μrad average horizontal steering corrector strength, the vertical correctors had a ring-wide average of about 16 μrad, and areas of the Tevatron had strengths of 70-90 μrad averaged over distances of 400 m or so. At 1 TeV, the maximum strength of a corrector is a little more than 100 μrad, so the available correction for general beam steering was limited in these locations. The interpretation of this effect was that these areas contained magnets which were systematically rolled toward the inside of the tunnel. This was verified by magnet roll angle measurements performed in October 2002 and January 2003.

The rolled dipoles and systematic corrections produce a “scalloped” vertical trajectory through the bending regions [3]. Although the distributed beam position monitors read zero displacements, the beam actually underwent ~0.5 mm excursions through these regions, assuming the magnets were rolled about the beam pipe axis. The fact that they are actually rolled about a different axis closer to the floor meant that the beam trajectory was closer to 1 mm or more from the center of the magnet coil.

Since the Tevatron dipoles have a sextupole component, a systematic vertical offset feeds down into a coupling term between the horizontal and vertical motion. The sextupole component is also known to vary as the logarithm of time due to persistent current effects at low magnetic fields, hence the coupling varies with time accordingly. This effect might explain some of the observed tune drift behavior during the Tevatron injection process, but was not nearly enough to explain the large skew quadrupole corrector settings.

In addition to the transverse coupling generated by orbit feed-down through rolled dipole magnets, the coupling due to observed quadrupole magnet rolls was calculated. Taken together, these effects were strong enough to explain a correction of the minimum difference between the two transverse tunes of amount $\Delta\nu \approx 0.03$. However, the setting required of the main skew quadrupole correction circuit to decouple the Tevatron was indicative of a minimum tune difference an order of magnitude larger.

## INJECTION EXPERIMENT

Beam studies were conducted to see if a local source of the coupling could be pin pointed. All of the skew quadrupoles were set to zero, and then beam position data were collected for the first several turns before dumping the beam. The beam was then mis-steered in the horizontal plane, and orbit data again collected on the first several turns. The orbit differences are shown in Figure 1. The purely horizontal error coupled fully into a vertical error in about 1.5 turns, and then back into the horizontal plane in another half period. [4]

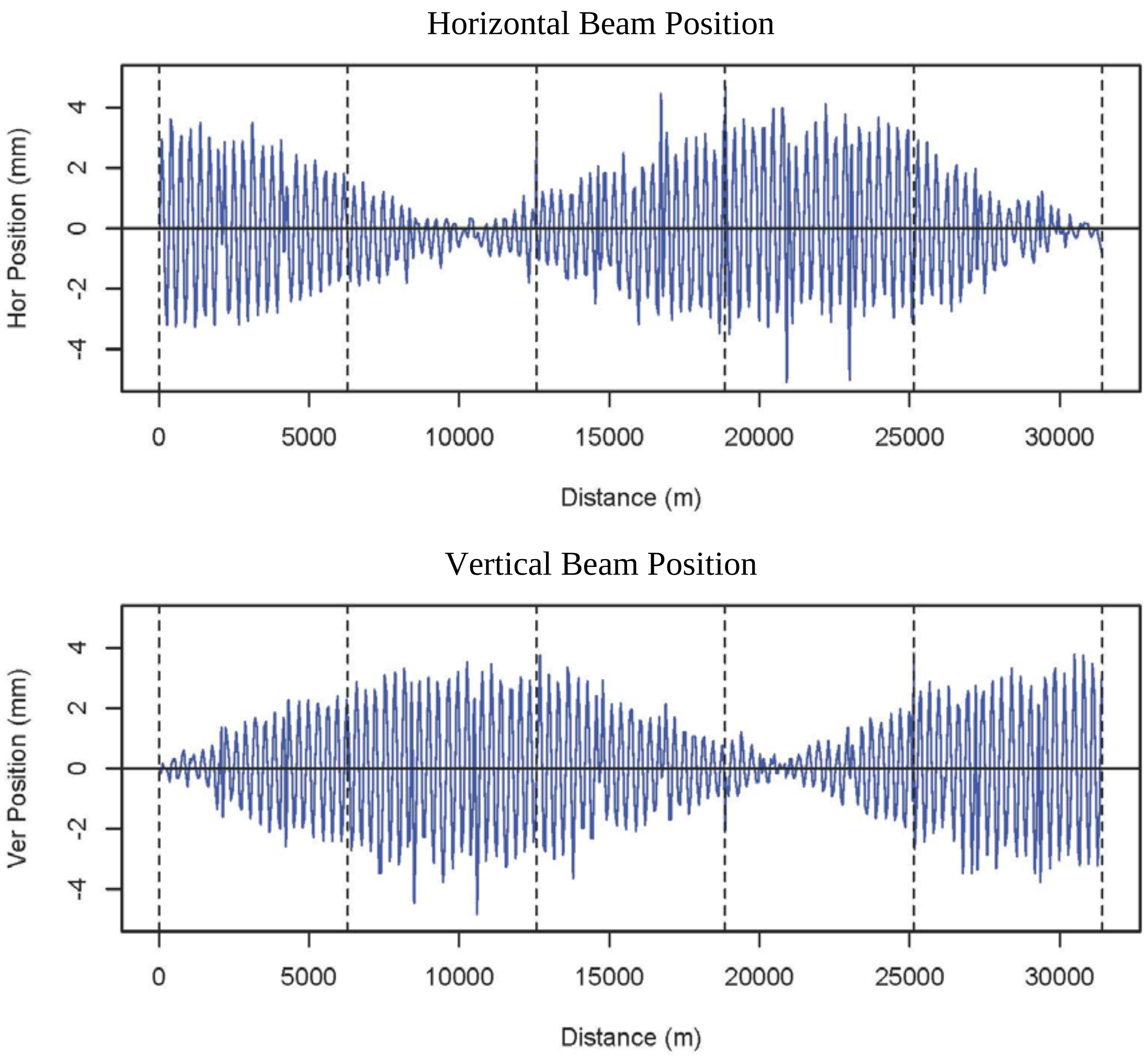

**Figure 1. An initial horizontal steering error couples completely to the vertical plane. This data set, taken in February 2003, had all skew quadrupole correctors turned off.**

The fact that the coupling builds up gradually and not at a few localized sources is indicative of a uniformly distributed source of skew quadrupole fields. The 3-turn period of the coupling is consistent with a tune split of order 0.3. It was quickly noted that a systematic skew quadrupole term, $a_1 \equiv (\partial B_x/\partial x)/B_0$, would account for this behavior, and would need to be of order $a_1 = 1.5 \times 10^{-4}$ inch$^{-1}$, or 1.5 "units." [5]

While regions of rolled dipoles existed at the time, the possibility that a systematic roll of the main quadrupoles could give rise to the coupling problem was unlikely, as the focusing and defocusing quadrupoles would have to be rolled in opposite directions. Direct roll measurements of the quadrupoles performed in 2002 also disproved this as the strong coupling source.

## MECHANICAL ISSUES

In January 2003, as plans were being made to correct the rolled Tevatron dipoles discussed above, the question arose of ensuring that the cryostat and coils moved with the warm iron magnet yoke. The hope was that the magnets could be realigned without disturbing the cryogenic and vacuum connections between adjacent magnets.

The Tevatron dipole cross section is shown in Figure 2. A warm iron yoke surrounds the cryostat tube and the collared coil within it. The collared coil is supported in the cryostat at nine stations down the 6-meter length of the magnet. At all but the center station the support is two rectangular blocks of G11, called “suspensions”, approximately 8.5 mm x 6.4 mm x 17 mm, aligned with each other, one between the outer cryostat skin and the heat intercept at liquid nitrogen temperature and the other between the heat shield and the phase single helium tube. These suspensions are arranged in grooves that allow the whole cryostat to move longitudinally relative to the yoke as it is cooled. At the middle station more robust G10 tubes, known as “anchors” support the cryostat, anchoring the center against both longitudinal motion and rotation relative to the yoke (a bane of earlier designs of the magnet).

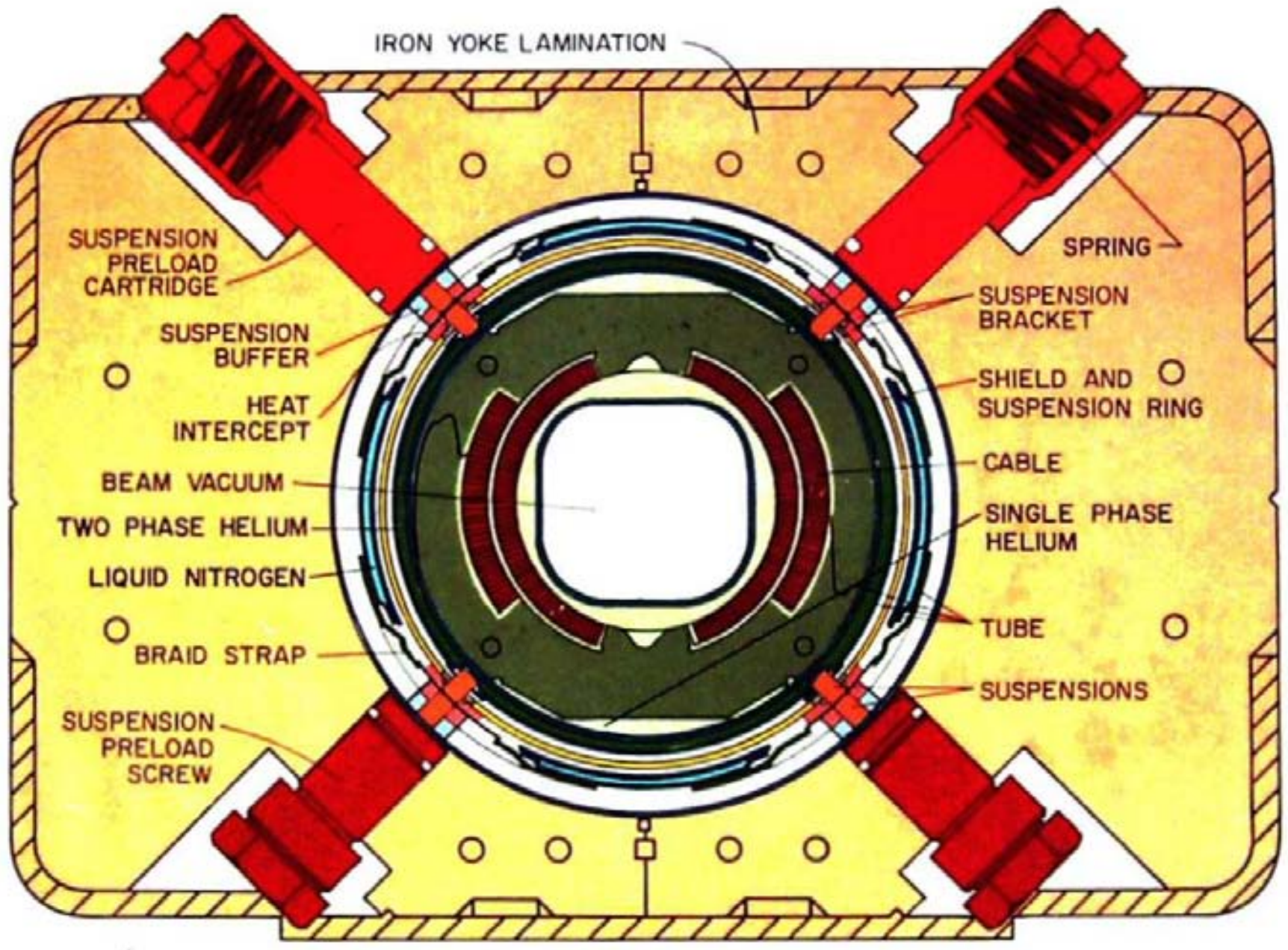

**Figure 2. Tevatron dipole cross section.**

At each station four supports are located at the 45 degree points. A screw or suspension cartridge through the yoke at each of these points contacts the outside of the cryostat. This allows the cryostat to rest stably in a cradle formed by the lower two screws ("bolts"), while being held down against those supports by two preloaded upper suspensions, spring-loaded cartridges known as "smart bolts." The pre-loaded springs maintain a force as the collared coil and cryostat shrink during cool down.

As discussed in more detail below, the magnets had been designed to allow adjustment of the coil position while the magnet was cold by adding and removing shims from the cryostat suspension system. During production this feature was used as a standard procedure to compensate for fabrication tolerances and mechanical differences between the top and bottom coils. A depth gauge can be used to measure the distance from the top of the cylinder to the top of the cartridge, a distance that became known as the "lift". For example, as the magnet was cooled, the collared coil shrank, the outer skin of the cryostat deformed slightly, the springs pushed the cylinders inward, and the lift increased. In 2002, lift measurements seemed to have potential for monitoring the movement of the cryostat relative to the yoke during magnet realignment.

To assess the viability of making these measurements in the tunnel and to check the stability of the cryostat over time, technicians measured 18 magnets during a day in February 2003 when the accelerator was down for maintenance. The original (paper) production records of the magnets were retrieved from off-site storage and the lift measurements extracted.

By the end of February the data had been analyzed and it was apparent that in those 18 magnets the coil had systematically dropped approximately 0.11 mm relative to the yoke. After a bit of thought it was realized that this was probably due to "creep," slow inelastic deformation under pressure, in the G11/G10 blocks that support the coil within the cryostat. The sizing of the G11 suspensions was a balance between mechanical strength and the heat load a larger conducting cross section would present. The stability of the blocks was the subject of a study at the time [6]. One participant remembers the 1980 guidance to "design the magnets to last for 20 years" [7].

The measured change was enough to produce a skew quadrupole component in the field of about 1.15 "units." Knowing that betatron coupling was a troubling issue in the Tevatron, but not knowing of the active calculations and measurements, one of the authors (Harding) sent e-mail to another (Syphers): "If every Tevatron dipole had developed with age a skew quadrupole component, how big would it need to be to explain the effects that are currently seen?" When the authors conferred and discovered that they were talking about numbers that matched quite well, they and the others involved agreed to report the results and propose further studies.

## UNDERSTANDING THE INADEQUACY OF THE CORRECTOR SYSTEM

Once the Tevatron dipole magnets were considered to be the main source of the coupling, further verifications were performed using beam measurements as well as magnet measurements. During Run I, upgrades of the low beta insertions required some of the existing spool pieces to be replaced with new devices. As a result, six skew quadrupoles were lost, one upstream and two downstream of each interaction point. Thus, while the distribution of $a_1$ was essentially uniform around the ring, its correction was no longer uniform.

The evenly spaced coupling errors from the dipole magnets were being compensated with correction circuits that now had gaps at two locations. Since the horizontal and vertical phases advance differently through a half cell, the main skew quad circuit was not able to fully correct the coupling caused by the distributed coupling errors. Powering the small skew quadrupole circuits near the straight sections in conjunction with the main circuit was necessary to fully minimize the global coupling. At injection, the most effective of the smaller circuits was the one surrounding the A0 straight section, and the skew quad at location A49 was also utilized. Since the smaller circuits consisted of only one or two elements, the strength of these correctors needed to be quite high to completely minimize the global coupling. Under these conditions, the non-uniformity of the coupling correction consisted not only of the gaps in the main circuit, but also the additional skew quadrupoles around A0 and at A49.

## VERTICAL DISPERSION

A horizontal offset in a skew quadrupole results in a vertical kick. A horizontal offset due to momentum in a skew quadrupole will therefore result in a vertical kick, creating vertical dispersion. Given the value of horizontal dispersion at the location of the skew quadrupole elements around A0 and at the gaps in the main skew quad circuit, a significant vertical dispersion resulted. The dispersion at A49 was very small so that a skew quadrupole there did not contribute to the vertical dispersion.

The dispersion can be measured easily by changing the RF frequency and recording the resulting orbit difference. Figure 3 shows the dispersion in both planes at the injection energy in early 2002.

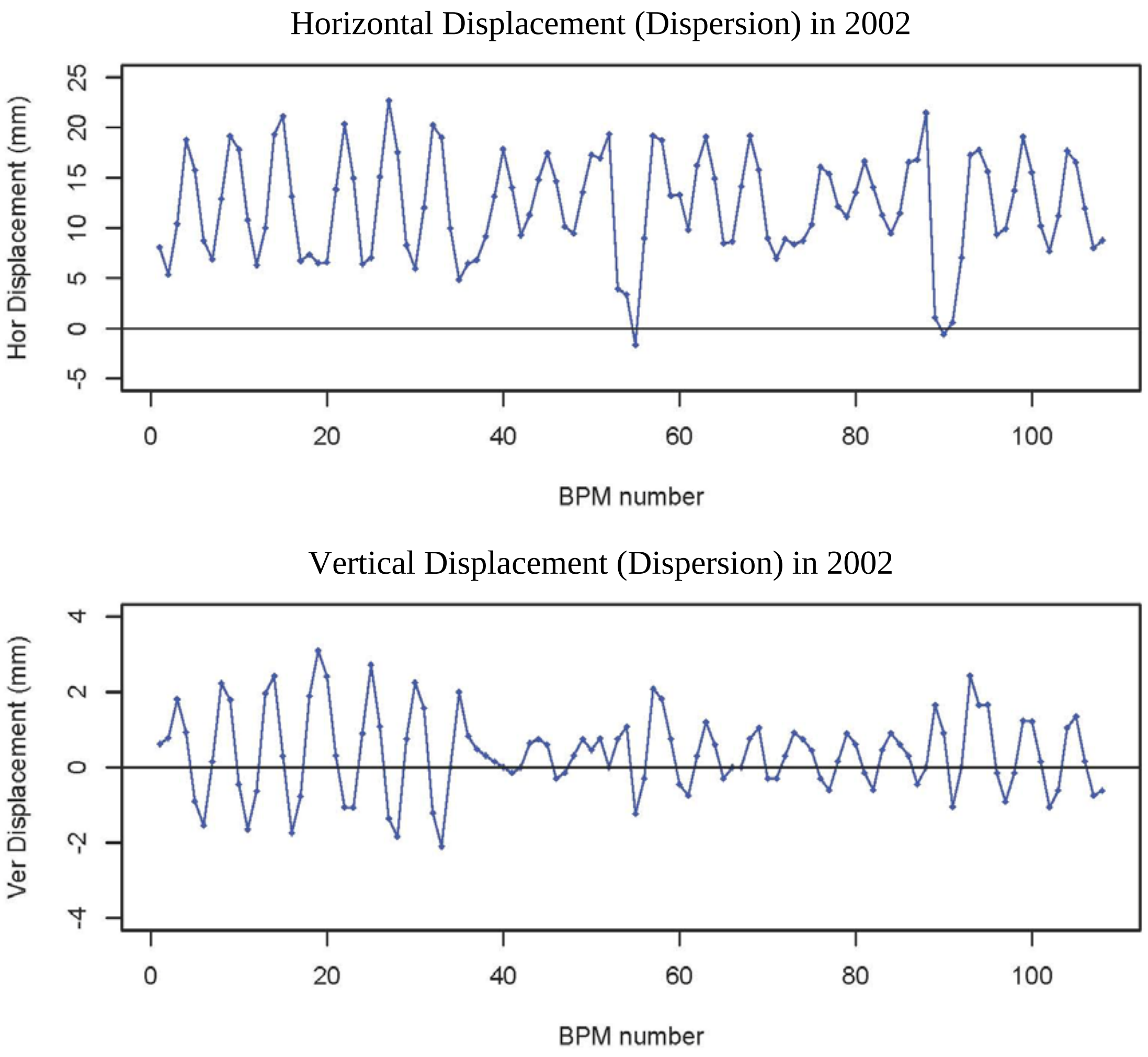

**Figure 3. The orbit difference given by an rf frequency shift, taken in early 2002. The position shift at a BPM is proportional to the dispersion at that location. The vertical displacements clearly show a coherent pattern.**

The value of the horizontal dispersion function oscillates in the arcs of the Tevatron between about 1 meter and 5 meters, and vanishes at the interaction points, in agreement with the design lattice. By design, the vertical dispersion should be zero, though a small "residual" vertical dispersion is always present due to various error sources, and would be expected to have a random distribution with peaks of approximately 0.2 m in the Tevatron. The vertical dispersion measurements, however, showed a fairly coherent dispersion wave of up to 0.8 meters that is largest between D0 and A0, smaller between B0 and D0, and very small between A0 and B0. This pattern is consistent with calculations based on coupling errors due to the strong correction circuits at A0 and A49, and the gaps in the correction circuit near B0 and D0. The vertical dispersion was near maximum value at the injection point at F0 leading to an emittance increase of 5 $\pi$ mm mrad or more for injection of coalesced Proton bunches.[8]

## MORE MAGNET DATA

The lifts of an additional sample of 66 magnets were measured in the tunnel during a subsequent short maintenance period, confirming the original distribution. Similar changes in the coil position were measured in spare dipoles that had not been in service. The original calculations from the production era were repeated, affirming the linear relationship between displacement and skew quadrupole error term, $a_1 = k \cdot y$, where k is 10.6 units/mm. Several spare dipoles were magnetically measured [9], confirming that the long term change in lift was reflected in the expected change in skew quadrupole. The amount of its life that a magnet had spent cold or warm did not seem to affect its creep.

Many more magnets have since been measured during the major shutdowns of 2003 and 2004. The distribution of lift change, averaged over each magnet, is plotted in Figure 4. The average shift of about 0.14 mm (0.0055”) is clearly seen, as is the breadth of the distribution. The difference between the average left and right lift changes is centered on zero, indicating no lateral movement.

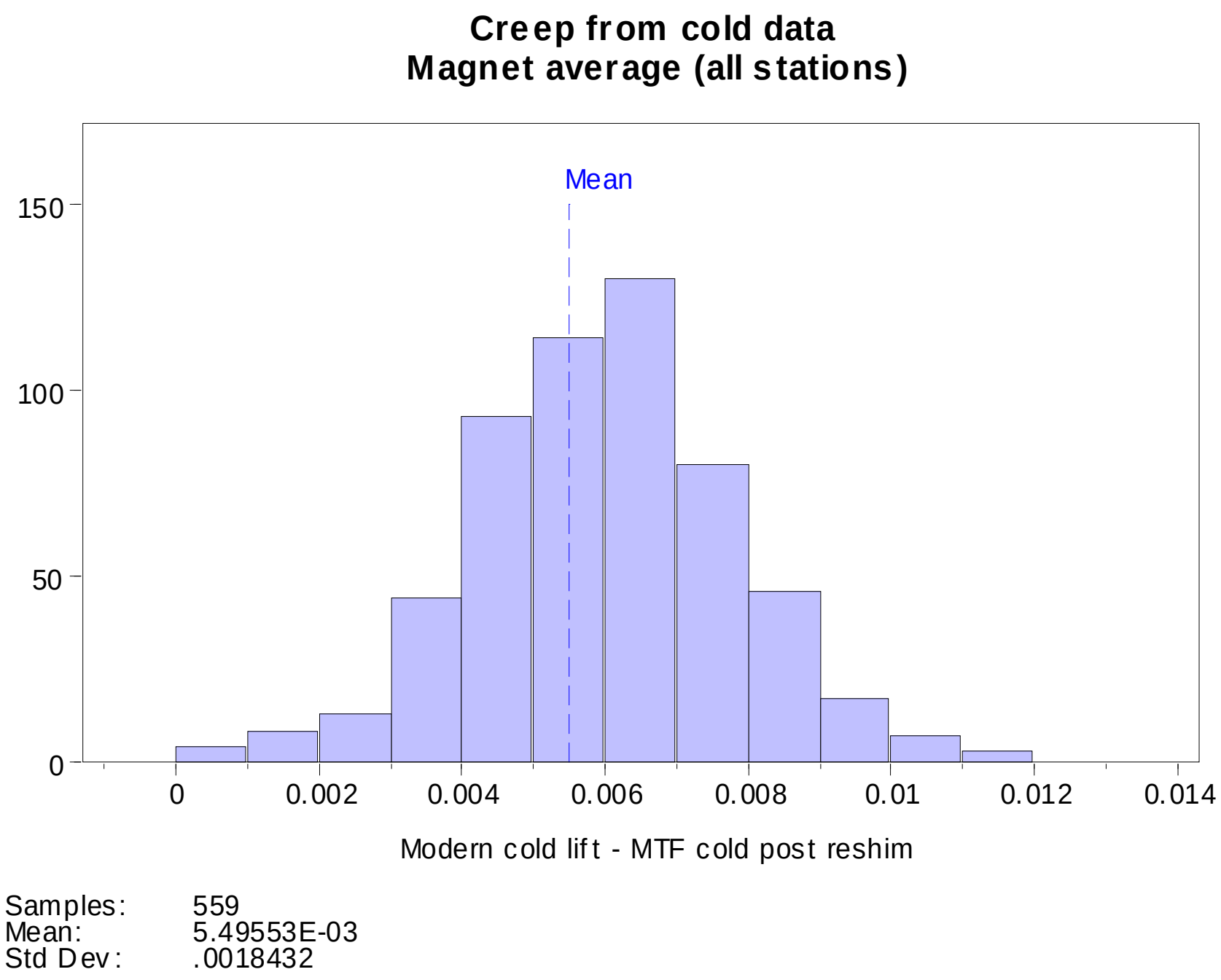

**Figure 4. Distribution of lift changes from production to now due to creep. For each magnet the change is averaged over all accessible stations.**

The spring force in the smart bolts is sufficiently greater than the weight of the cryostat that we take the compressive force on the upper and lower suspensions to be equal and interpret the change in lift to be creep equally distributed between the upper and lower sets. Note also that an equal creep all around results in a change in the coil position that is the vector sum of the creep in the two lower suspensions.

## MAGNET FIELD CORRECTION

When there is any asymmetry in the coils of a conductor-dominated dipole magnet, an error field is produced, which we characterize by the coefficients of its harmonic decomposition. For convenience we normalize the error field to the central field, take the reference radius to be 25.4 mm, and quote the coefficients in "units" of parts in $10^4$. This means that if we can bring the coefficients into the range of unity, the magnetic field will be acceptable over the useful aperture of the magnet.

The asymmetry may be in the geometry of the coils, in their placement in the yoke, or in the superconducting cable. Of particular concern here is the fact that a normal (skew) quadrupole term is produced by the horizontal (vertical) displacement of the coils from the center with a linear dependence. By adjusting the cryostat and collared coil position in the yoke, the quadrupole components of the field could be eliminated or cancelled during production.

During assembly brass shims were placed on the end of all four bolts at each station to roughly center the coil in the yoke. Each of the almost 1000 (including spares) production magnets was measured in detail.[10] After quench testing, the harmonic components of the magnet field integral were measured at one current and the magnet was shimmed according to the guidance provided by a computer calculation. The cryostat support system described above allowed these adjustments while the magnet was cold. While the magnet was still on the test stand, one opposing pair bolts were removed at a time, shims were shifted from one bolt to the other to reposition the coil, and the bolts screwed back in. The magnet was then measured at multiple currents on both the up and down ramps. This data set includes every magnet and provides a measurement of the entire magnet field integral. The specification called for both quadrupole components to be adjusted, if needed, to be within one unit of zero. If the first try did not succeed, a new shim adjustment could be calculated and the process repeated. Figure 5 shows the final distribution of the skew quadrupole component.

## MAGNETIC STUDIES

When the creep issue was recognized, a program of measurements was under way at the Fermilab Magnet Test Facility (MTF) studying the dynamic behavior of Tevatron dipoles.[11] The work was expanded to include some measurements related to the skew quadrupole questions. Using modern techniques, these measurements can capture the field during a ramp at very densely spaced currents. Only a small sample of spare magnets was available for measurement, however, not necessarily representative of the installed magnets. The modern probe is short (0.83 m) and the fixturing was inadequate to allow a reliable full length measurement to be assembled.

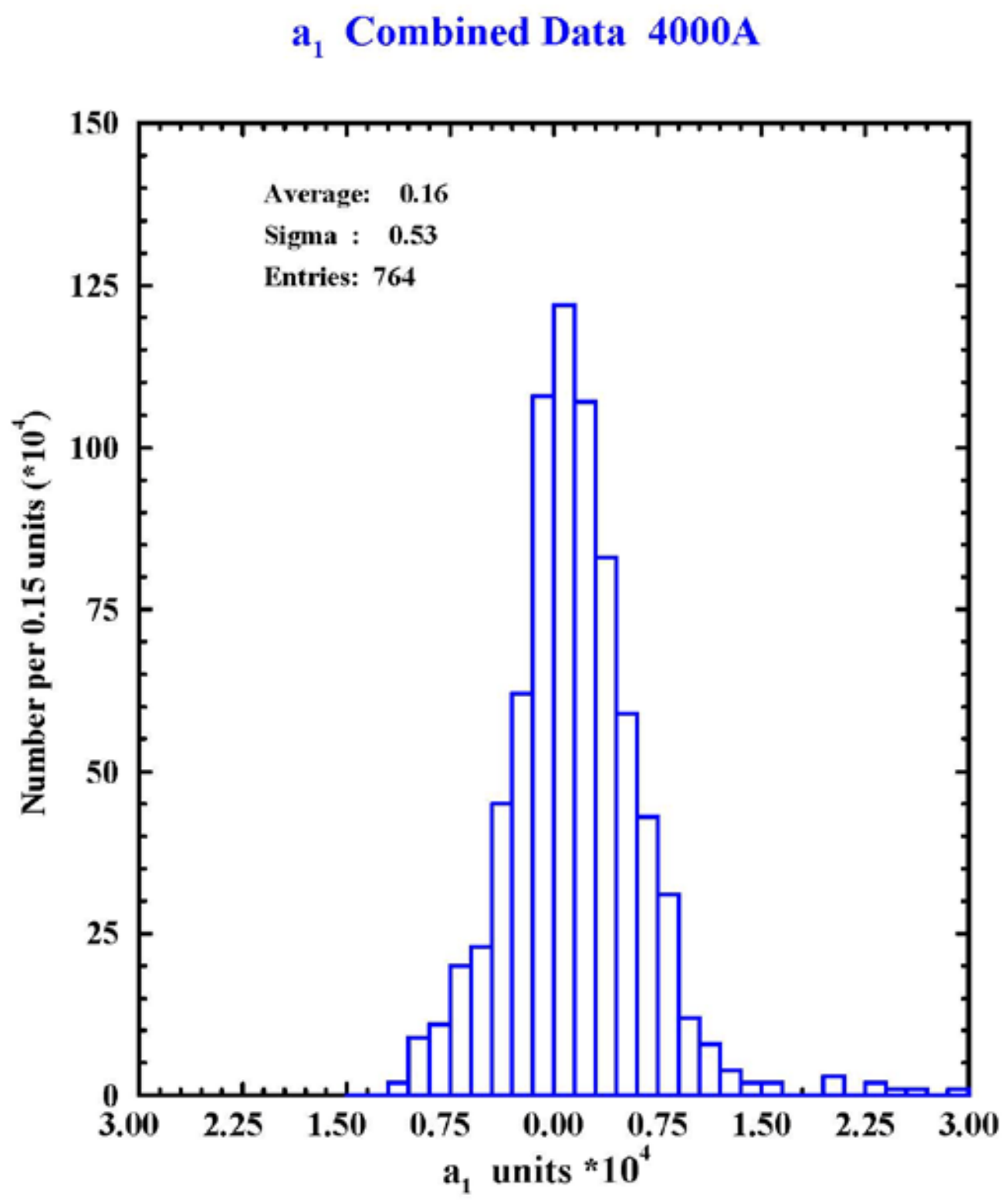

**Figure 5. Final distribution of skew quadrupole in Tevatron dipoles after production shimming.**

*Incomplete Reshimming*

Recognizing that the production experience in shimming close to a thousand magnets had validated the technique exhaustively, only a few people insisted on seeing another demonstration of the basic procedure. There were, however, a few more subtle questions. As part of the preparation for reshimming magnets in the tunnel, one dipole was reshimmed on the test stand at MTF while cold, with harmonic measurements before and after each step.

Since it was anticipated that on many magnets one or more stations would be inaccessible due to various interferences in the tunnel, especially at the magnet ends, it was important to establish the behavior after non-uniform reshimming. The short probe of the modern system allowed measurement of the field as a function of position. Artifacts such as effects of the superconducting cable pitch confused the interpretation of these measurements, but in the end it was agreed that as many as two stations could be left unshimmed without a significant impact on the integral.

*Skew Quadrupole Variation with Excitation*

Archival data showed that the skew quadrupole contribution of many magnets varied as a function of current, suggesting coil movement under excitation. (See Figure 6.) The effect was small and there were magnets that exhibited both signs of variation.

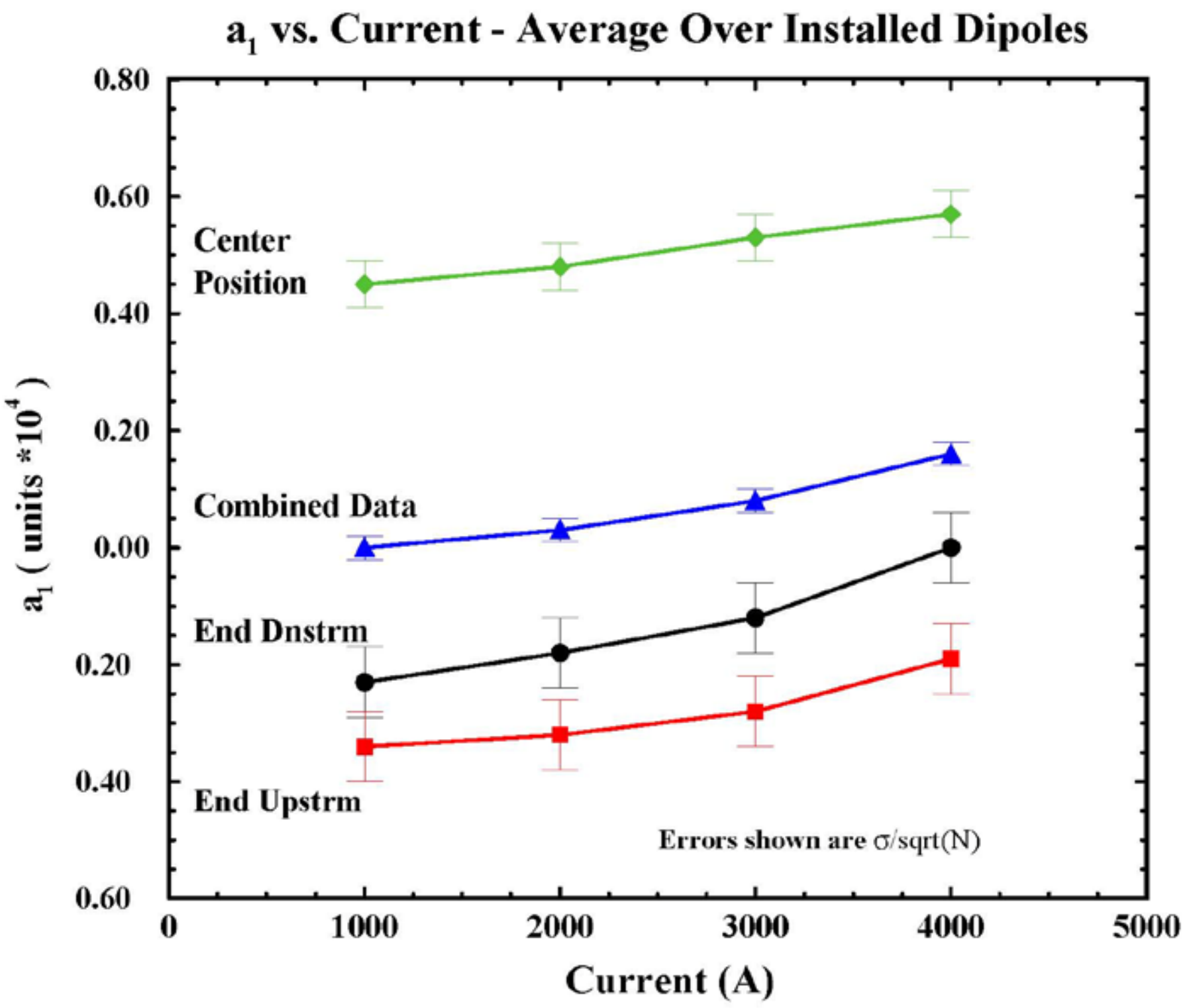

**Figure 6. Dependence on magnet current of the average skew quadrupole, measured for each third of the a dipole magnets and over the whole magnets ("combined").**

One possible mechanism considered was that if the coil was too high, the decentering Lorentz force might overcome both gravity and the springs of the smart bolts. Before this phenomenon was recognized during production, two magnets were damaged. After a limit was put on the allowed coil position there was no further damage, but small movements were not ruled out. A dedicated test was devised with the modern MTF test to study possible coil motion. First, gross motion was checked by measuring lifts during a full current ramp cycle. Then the coil was purposely positioned up, down, left, and right by far more than normal, exaggerating any decentering forces. Under those conditions no changes were observed in the quadrupole field as a function of the decentering force, as shown in Figure 7. If there were any coil motion, it would be exaggerated in opposite directions with the coil high and low. This is not observed in the measurement.

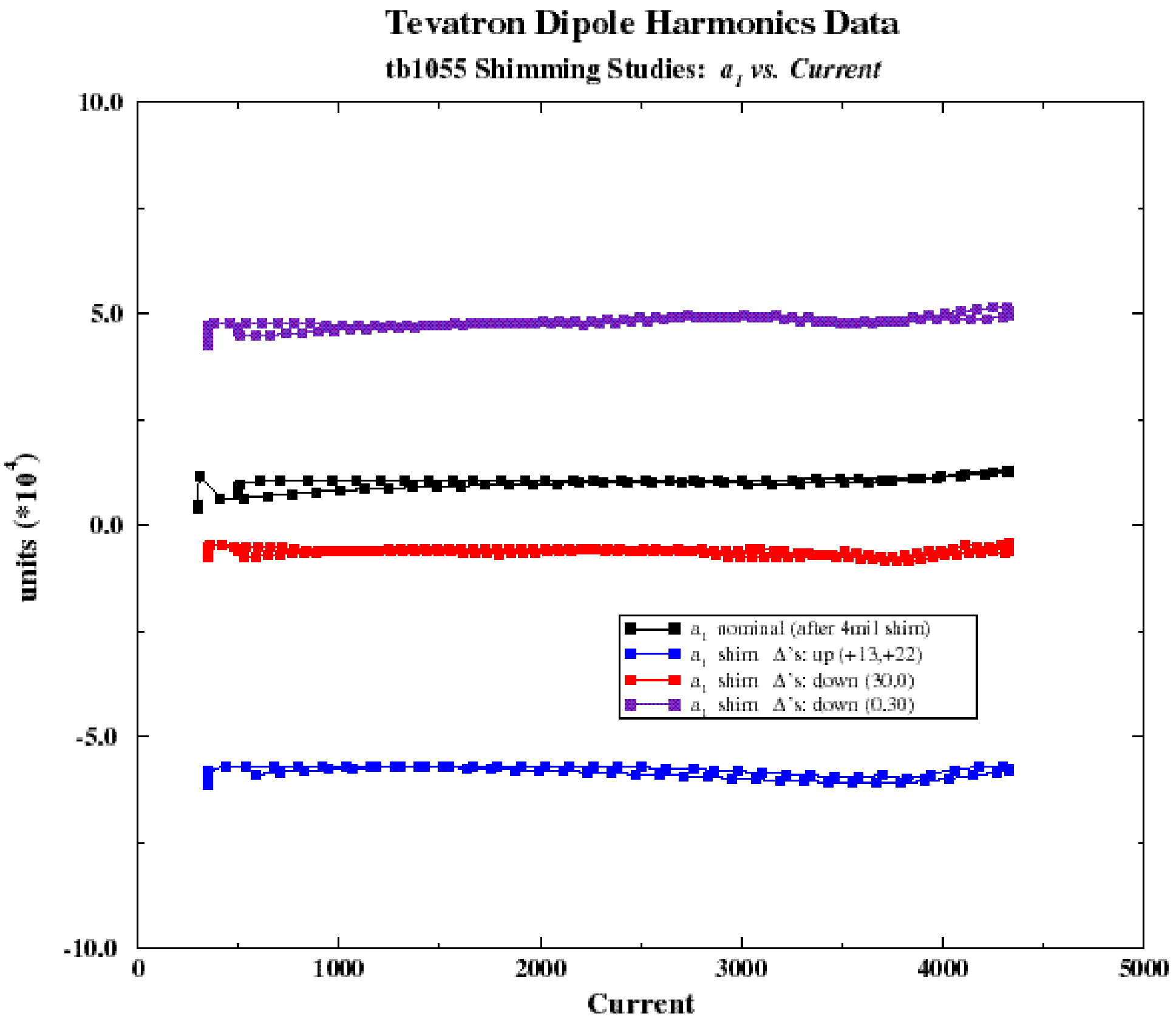

**Figure 7. Skew quadrupole as a function of current for one magnet with the coil offset from center.**

## THE RESHIMMING PLAN

During production the results of shimming could be verified immediately. This would not be possible during reshimming in the tunnel. Examination of the production lift data quickly led us to the conclusion that it was not sufficiently reliable on a magnet-by-magnet basis to warrant the effort of individual treatment. Measuring the lifts of each magnet, comparing them with the production values, and reshimming accordingly would have been fraught with possibilities for error. Instead, it was decided to treat all magnets as though they had changed by the same amount. The addition of a standard shim 3 mil (0.076mm) was agreed upon, raising the coils by 2.12 mil (0.054 mm).

With 774 dipole magnets in the Tevatron, and with 18 shim locations per magnet, it was clear that to fix the problem completely would take significant tunnel access time. Trial runs above ground had shown that although the job in the tunnel was tedious, requiring the technicians to crawl over the magnets in tight quarters, a crew of three could comfortably reshim two magnets per day. It was not clear, however, how many exceptional cases would arise requiring extra time and how well the crews would hold up with time, even with careful ergonomic planning. So, given some uncertainty in the effectiveness of the reshimming operation, the limited resources, and the uncertainty in the amount of time to reshim the magnets, we decided to start with a modest correction program. We initially focused on the region where the regularity of skew quad correction system had been interrupted, with the idea that the balance of the ring could be reshimmed in future maintenance periods.

For the first run in the fall of 2003 we chose to reshim 106 magnets divided between the two sides of each intersection region, the areas that had had their skew correction systems disrupted. After this work, the accelerator performance was significantly improved, so another 12 magnets were reshimmed in March 2004 during a one-week shutdown. The next major period of repair was the following fall (September-November 2004) when 412 dipoles were re-shimmed. This set of dipoles consisted of magnets that were at least 4 dipole locations from the nearest main skew quadrupole corrector. The performance continued to improve after the second repair, and the remaining 244 dipoles were adjusted in early (March-April) 2006.

The sensitivity of the operation led us to augment our usual quality assurance procedures with more formal analyses.[12] The sheer volume of measurements, 18 lift measurements for each magnet before and after the reshimming, required extensive above ground monitoring in addition to the teams of technicians in the tunnel.

## ACCELERATOR PERFORMANCE IMPROVEMENTS

### *Reduced Skew Quadrupole Correction*

After each round of reshimming the required current in the main skew quadrupole correction circuit decreased, giving clear indication that the reshimming was reducing the skew quadrupole component in the dipoles. Note that in Figure 8 the current is negative and zero is at the top of the plot. There is only a small change after the first stage, which only addressed the regions where the main skew quadrupoles had been removed. The two stages that addressed the regions covered by the main skew quad circuit were directly effective. The current on the injection front porch has decreased by a factor of six. All other skew quad circuits were reduced to near zero at the injection energy by the first stage of reshimming, as expected. Stronger coupling corrections are needed at high energy.

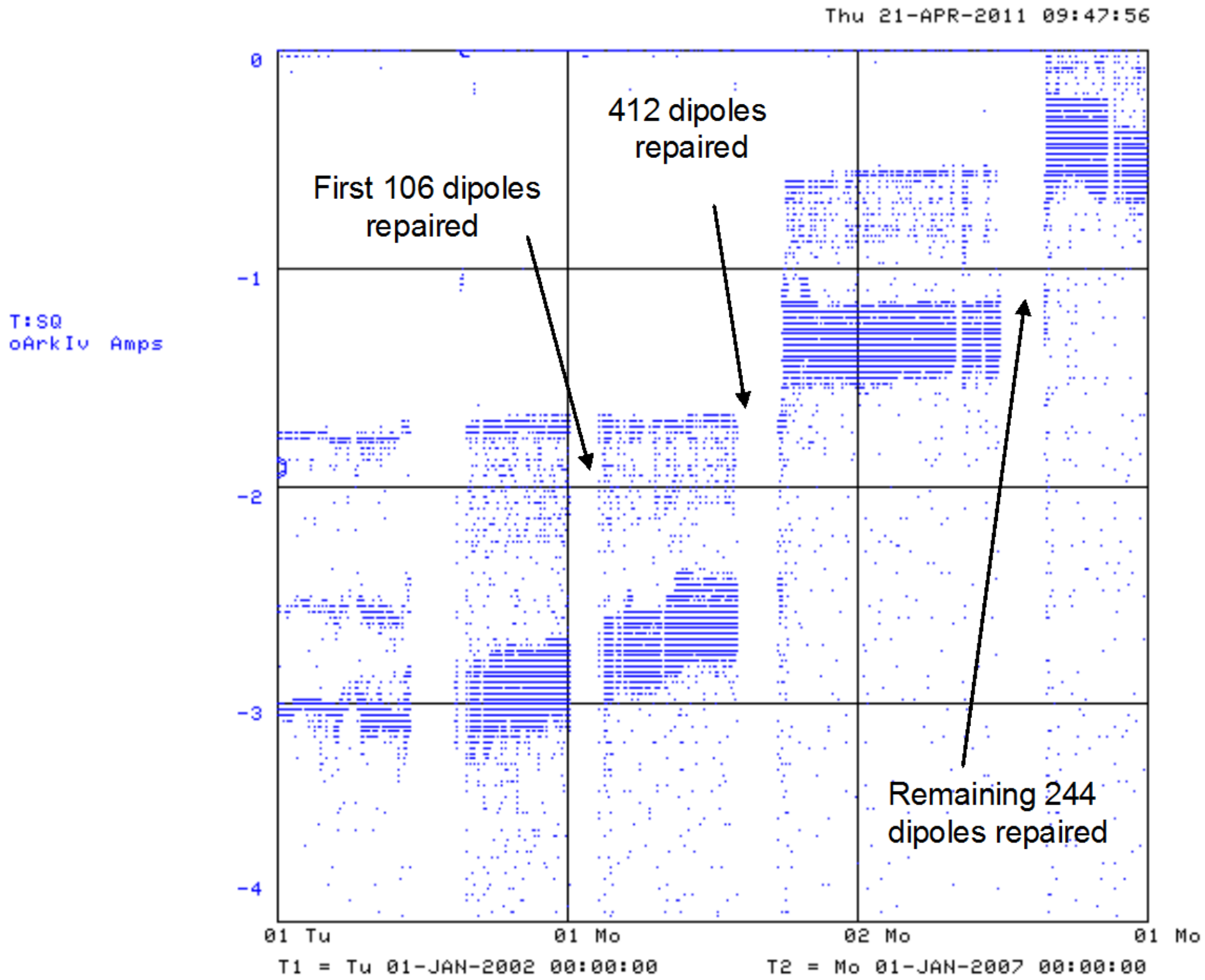

**Figure 8. Six year plot of current in main skew quadrupole circuit. The darkest bands show the current during the time on the injection porch.**

*Reduced Vertical Dispersion*

The vertical dispersion has been reduced with each phase of the reshimming. A dispersion measurement from 2011, shown in Figure 9, can be compared with the starting dispersion measurement from 2003, shown in Figure 2. The 2011 measurement shows vertical closed obit changes of about 1 mm, oscillating with roughly constant amplitude, where in 2003 the amplitudes varied around the ring, peaking at over three times this amount.

Horizontal Displacement (Dispersion) in 2011

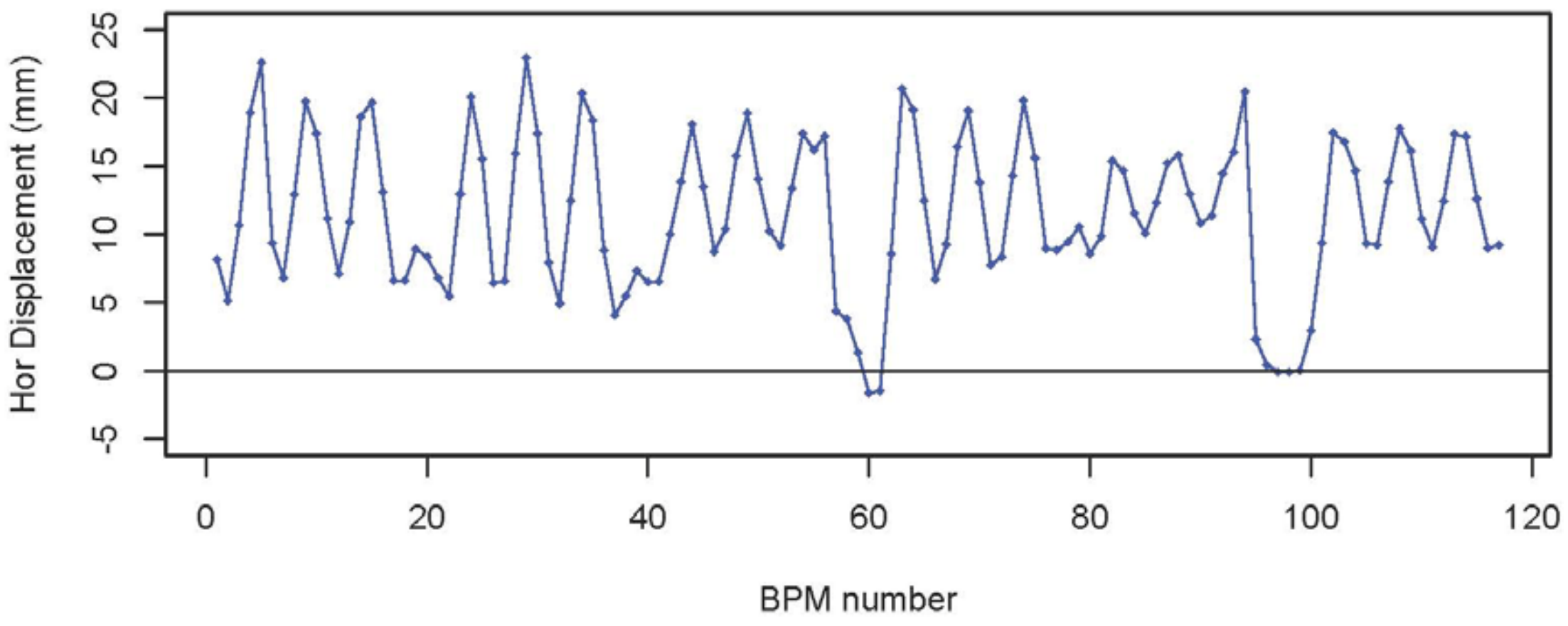

Vertical Displacement (Dispersion) in 2011

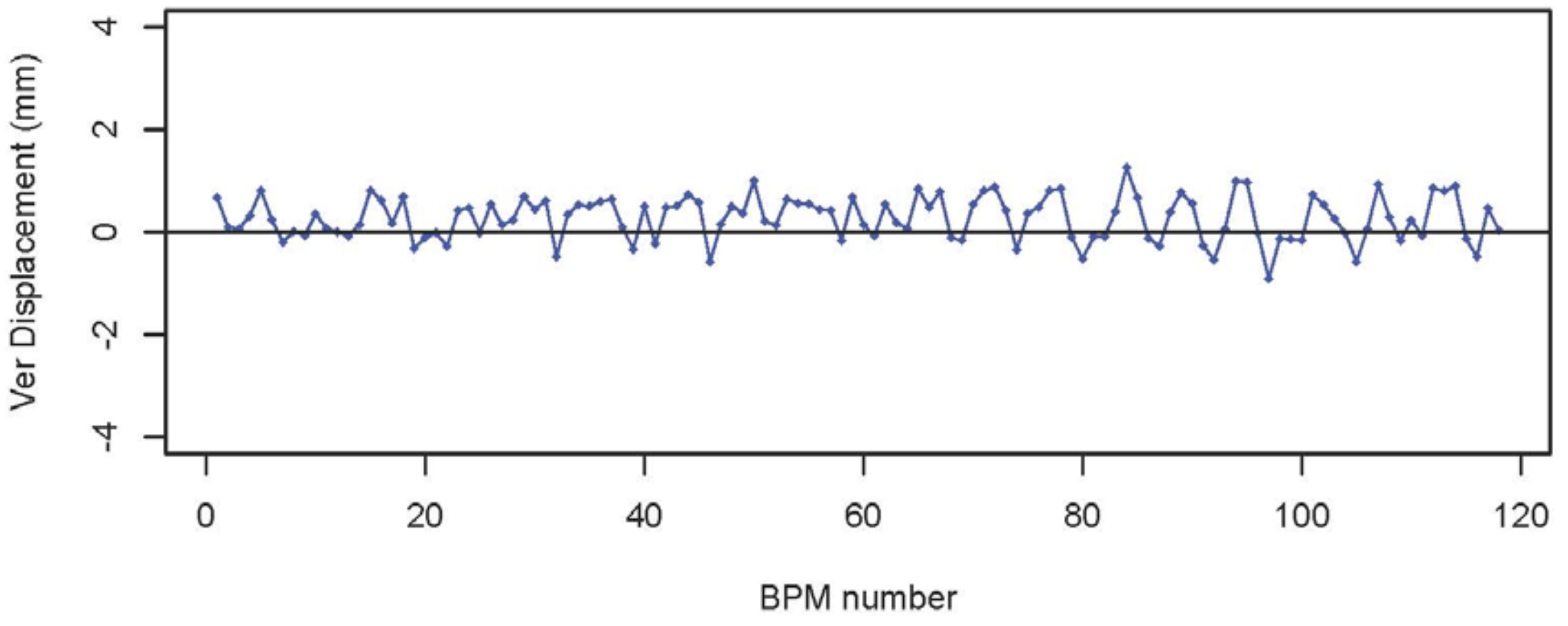

**Figure 9. Results of dispersion measurement made in 2011 to be compared with the 2002 measurement shown in Figure 2.**

*Reduced Emittance Growth at Injection*

Most importantly, the emittance growth at injection due to the mismatch between the optics of the rings and the injection lines has also decreased dramatically. Figure 10 shows the proton injection emittance growth by store number spanning the first two major reshimming periods. Where in early 2003 the emittance growth was typically over 7π mm-mr, immediately after the first round of reshimming the emittance growth dropped to 4π mm-mr. The second major round of reshimming reduced the growth by another 2π mm-mr.

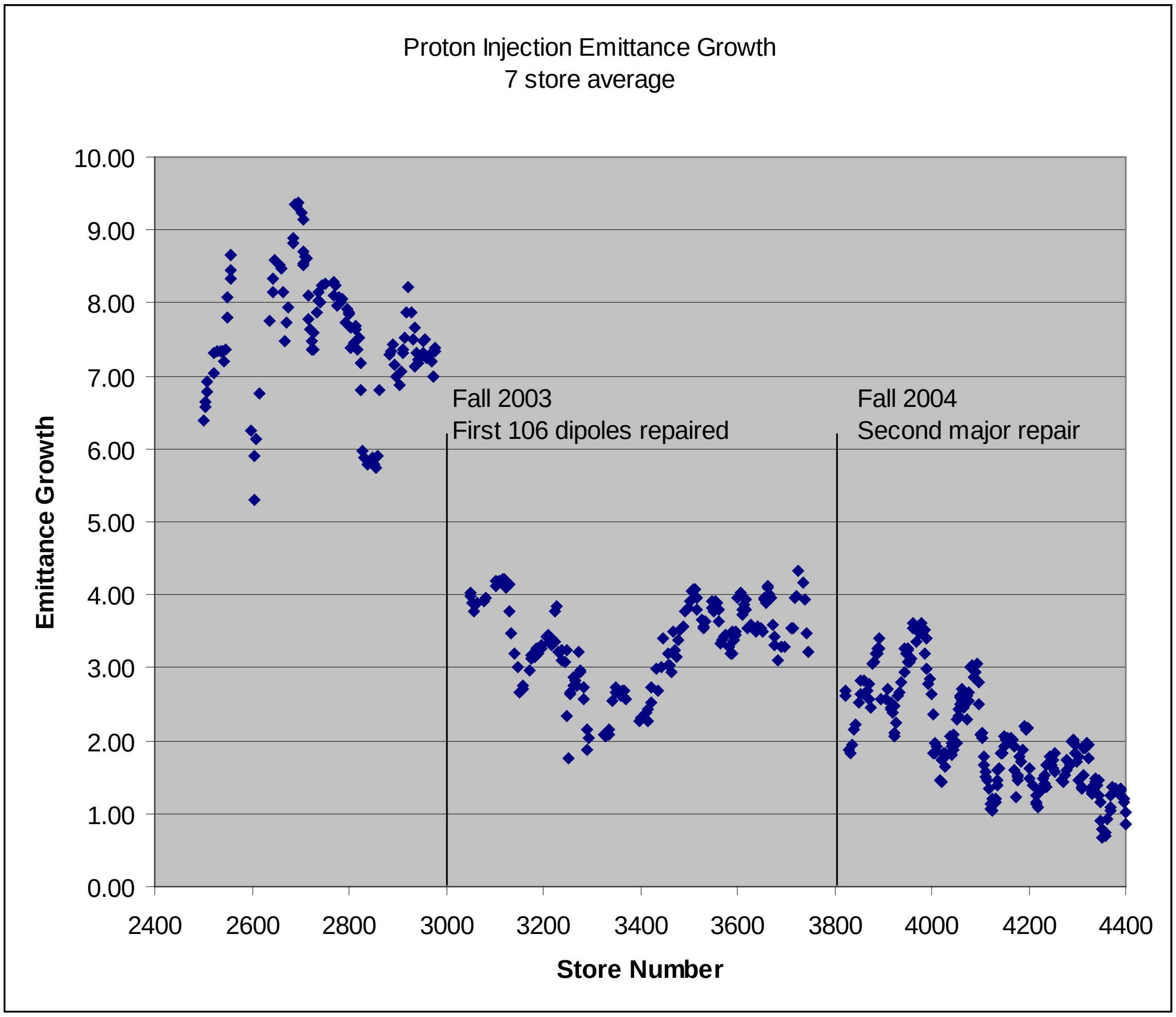

**Figure 9. Seven-store average of the Proton injection emittance growth spanning the period of the first two major repair periods.**

*Indirect Benefits*

The reshimming has had an indirect benefit as well. With the major sources of coupling removed, other sources (such as two rolled arc quadrupoles) could be identified and the entire Tevatron lattice restored closer to its design. Correspondingly beam measurements have become much cleaner and easier to interpret, greatly enhancing operation of the collider.

**SUMMARY**

Many issues plagued the ramp-up of luminosity during Run II of the Tevatron collider program. While adjustments of the global coupling of the accelerator using a systematic skew quadrupole magnet circuit could allow the operators to place the betatron tunes at desirable settings, other confusion ensued. For example, horizontal corrections that seemed to induce vertical oscillations hampered the commissioning of beam damper systems, and unexplained emittance growth – particularly in the vertical degree of freedom -- upon injection was difficult to track down. The required settings of the quadrupole corrector circuit indicated that large sources of coupling were present – and beam measurements soon verified that the source was everywhere. Independently but in parallel, the explanation emerged from both the interpretation of beam measurement data and of magnet measurement data taken in the tunnel – the Tevatron dipole magnets were the source and a mitigation scheme was required. The coils of essentially every dipole magnet in the tunnel were shimmed to re-center them in their iron yokes. The process, which took careful planning and special training, was strategically implemented over a period of years during planned shutdowns of the collider program, restoring the Tevatron to its 1980's state of small residual coupling, both globally and locally.